\documentclass[preprint,12pt]{elsarticle}
\usepackage{graphicx}
\usepackage{amssymb}
\usepackage{amsmath}

\journal{Physics Letters B}

\begin{document} 

\begin{frontmatter} 

\title{The pion-nucleon $\sigma$ term from pionic atoms} 

\author{E.~Friedman}
\author{A.~Gal\corref{cor1}}
\cortext[cor1]{Corresponding author: Avraham Gal, avragal@savion.huji.ac.il}
\address{Racah Institute of Physics, The Hebrew University, 91904 Jerusalem, 
Israel}

\begin{abstract} 
Earlier work suggested that the in-medium $\pi N$ threshold isovector 
amplitude $b_1(\rho)$ gets renormalized in pionic atoms by $\sim 30\%$ away 
from its $\rho=0$ free-space value, relating such renormalization to the 
leading low-density decrease of the in-medium quark condensate $<\bar q q>$ 
and the pion decay constant $f_{\pi}$ in terms of the pion-nucleon $\sigma$ 
term $\sigma_{\pi N}$. Accepting the validity of this approach, we extracted 
$\sigma_{\pi N}$ from a large-scale fit of pionic-atom level shift and width 
data across the periodic table. Our fitted value $\sigma_{\pi N}=57\pm 7
$~MeV is robust with respect to variation of $\pi N$ interaction terms other 
than the isovector $s$-wave term with which $\sigma_{\pi N}$ was associated. 
Higher order corrections to the leading order in density involve some 
cancellations, suggesting thereby only a few percent overall systematic 
uncertainty. The value of $\sigma_{\pi N}$ derived here agrees with 
values obtained in several recent studies based on near-threshold 
$\pi N$ phenomenology, but sharply disagrees with values obtained 
in recent direct lattice QCD calculations. 
\end{abstract} 


\end{frontmatter}

\section{Introduction}
\label{sec:intro}

The $\pi N$ $\sigma$ term 
\begin{equation} 
\sigma_{\pi N}=\frac{{\bar m}_q}{2m_N}\sum_{u,d} \langle N|{\bar q}q|N\rangle, 
\,\,\,\,\,\, {\bar m}_q =\frac{1}{2}(m_u + m_d), 
\label{eq:sigdef} 
\end{equation} 
records the contribution of explicit chiral symmetry breaking to the nucleon 
mass $m_N$ arising from the non-zero value of the $u$ and $d$ quark masses 
in QCD. A wide spectrum of evaluated $\sigma_{\pi N}$ values, from about 
20 to 80~MeV, was compiled by Sainio back in 2002~\cite{Sainio02}. Recent 
evaluations roughly fall into two classes: (i) pion-nucleon low-energy 
phenomenology, using $\pi N$ $s$-wave scattering lengths derived precisely 
from pionic hydrogen and deuterium, results in calculated values of 
$\sigma_{\pi N}\sim (50-60)$~MeV \cite{AMO12,CYZ13,Hof15,DCH16,RdE18}, 
the most recent of which is 58$\pm$5~MeV, whereas (ii) recent lattice 
QCD calculations reach values of $\sigma_{\pi N}\sim (30-50)$~MeV 
\cite{Hor12,BMW16,chiQCD16,ETM16,RQCD16,JLQCD18}, the most recent 
of which is 26$\pm$7~MeV. However, when augmented by chiral perturbation 
expansions such lattice calculations may lead also to values of about 
50~MeV, see e.g. Refs.~\cite{LTW00,ALCV13,RGM15,RLG18}. This spread of 
calculated $\sigma_{\pi N}$ values is discussed further in the concluding 
section.{\footnote{Some works denote the entity defined by the r.h.s. of 
(\ref{eq:sigdef}) as the nucleon $\sigma$ term $\sigma_N$. The notation 
adopted in the present work, $\sigma_{\pi N}$, follows that of recent 
works rooted in the low-energy non-strange sector of hadronic physics, 
e.g.~\cite{Hof15}.}} 

Here we show that the wealth of data on pionic atoms across the periodic 
table provides a precise determination of $\sigma_{\pi N}$. The experimental 
database for pionic atoms is the most extensive of all hadronic atoms 
\cite{BFG97,FGa07}, offering a useful test-ground for studying in-medium 
effects. On the theory side, the near-threshold pion-nucleus optical 
potential $V_{\rm opt}$ is given by single-nucleon $\pi N$ interaction terms 
approximated by their free-space values, with relatively small contributions 
from absorption on two nucleons. Our recent analysis of pionic atoms 
\cite{FGa14} demonstrated robustness in the quality of fitting the data 
against details of the applied analysis methodology. 

The starting point in discussing in-medium renormalization in pionic 
atoms is that the free-space isoscalar and isovector $\pi N$ scattering 
lengths derived in a chiral perturbation calculation~\cite{Baru11} 
from pionic hydrogen and pionic deuterium precise X-ray 
measurements~\cite{Hennebach14,Strauch11}, 
\begin{equation}  
b_0^{\rm free}=0.0076\pm 0.0031\;m_{\pi}^{-1}, \;\;\;\;\; 
b_1^{\rm free}=-0.0861\pm 0.0009\;m_{\pi}^{-1}, 
\label{eq:Baru} 
\end{equation} 
are well approximated by the Tomozawa-Weinberg (TW) leading-order chiral 
limit~\cite{TWe66}
\begin{equation} 
b_0^{\rm TW}=0,\;\;\;\;\; b_1^{\rm TW}=-\frac{\mu_{\pi N}}{8\pi f^{2}_{\pi}}
=-0.079\;m_{\pi}^{-1},  
\label{eq:TW} 
\end{equation} 
\noindent 
where $\mu_{\pi N}$ is the pion-nucleon reduced mass and $f_{\pi}=92.2$~MeV 
is the free-space pion decay constant. This expression for the isovector 
amplitude $b_1$ suggests that its in-medium renormalization is directly 
connected to that of the pion decay constant $f_{\pi}$, given to first 
order in the nuclear density $\rho$ by the Gell-Mann - Oakes - Renner (GOR) 
expression \cite{GOR68} 
\begin{equation} 
\frac{f_\pi^2(\rho)}{f_\pi^2} = \frac{<\bar q q>_{\rho}}{<\bar q q>} 
\simeq 1 - \frac{\sigma_{\pi N}}{m_{\pi}^2 f_{\pi}^2}\,\rho, 
\label{eq:fpi} 
\end{equation} 
\noindent 
where $<\bar q q>_{\rho}$ stands for the in-medium quark condensate 
and $\sigma_{\pi N}$ is the pion-nucleon $\sigma$ term. The decrease of 
$<\bar q q>_{\rho}$ with density in Eq.~(\ref{eq:fpi}) marks the leading 
low-density behavior of the order parameter of the spontaneously broken 
chiral symmetry. Recalling the $f_{\pi}$ dependence of $b_1^{\rm TW}$ 
in Eq.~(\ref{eq:TW}), Eq.~(\ref{eq:fpi}) suggests the following density 
dependence for the in-medium $b_1$:  
\begin{equation} 
b_1=b_1^{\rm free}\left(1-\frac{\sigma_{\pi N}}{m_{\pi}^2f_{\pi}^2}
\rho\right)^{-1}.  
\label{eq:ddb1} 
\end{equation} 
\noindent
In this model, introduced by Weise \cite{Wei00,Wei01}, the explicitly 
density-dependent $b_{1}(\rho)$ of Eq.~(\ref{eq:ddb1}) figures directly 
in the pion-nucleus $s$-wave near-threshold potential. Studies of pionic 
atoms~\cite{KYa01,Fri02,GGG02,YHi03,KKW03,KKW03c,FGa03c,FGa03,FGa04,Suz04} 
and low-energy pion-nucleus scattering~\cite{Fri04,Fri05} confirmed that 
the $\pi N$ isovector $s$-wave interaction term is indeed renormalized in 
agreement with Eq.~(\ref{eq:ddb1}). It is this in-medium renormalization 
that brings in $\sigma_{\pi N}$ to the interpretation of pionic-atom data. 
However, the value of $\sigma_{\pi N}$ was held fixed around 50 MeV in these 
studies, with no attempt to determine its optimal value. 

In the present work, we kept to the $\pi N$ isovector $s$-wave amplitude 
$b_1$ renormalization given by Eq.~(\ref{eq:ddb1}), but adopted a reversed 
approach of fitting $\sigma_{\pi N}$ to a comprehensive set of pionic atoms 
data across the periodic table. Other real $\pi N$ interaction parameters 
fitted simultaneously with $\sigma_{\pi N}$ converged at expected free-space 
values. Holding these parameters fixed at the converged values, except for 
the tiny isoscalar $s$-wave amplitude $b_0$ which is renormalized primarily 
by a double-scattering term (see below), we get a best-fit value of 
$\sigma_{\pi N}=57\pm 7$~MeV. 

The paper is organized as follows. In Sect.~\ref{sec:meth} we outline 
the methodology applied to fitting pionic atoms data. Results are given 
in Sect.~\ref{sec:res}, followed by discussion in Sect.~\ref{sec:disc} 
of estimated deviations from the linear-density expression (\ref{eq:fpi}) 
and their impact on the value derived for $\sigma_{\pi N}$.

\section{Methodology} 
\label{sec:meth} 

Here we briefly review the methodology applied in our recent work~\cite{FGa14} 
to dealing with pionic atoms data, using energy-dependent optical potentials 
within a suitably constructed subthreshold model. For a recent review focusing 
on $K^-$ and $\eta$ nuclear near-threshold physics, see Ref.~\cite{GFB14}. 
The pion self-energy operator $\Pi(E,\vec p,\rho)$ in nuclear matter 
of density $\rho$ enters the in-medium pion dispersion relation~\cite{FGa07} 
\begin{equation} 
E^2-{\vec p}^{~2}-m_{\pi}^2-\Pi(E,\vec p,\rho)=0, 
\label{eq:disp} 
\end{equation} 
\noindent 
where ${\vec p}$ and $E$ are the pion momentum and energy, respectively, 
in nuclear matter of density $\rho$. The resulting pion-nuclear optical 
potential $V_{\rm opt}$, defined by $\Pi(E,\vec p,\rho)=2EV_{\rm opt}$, 
enters the wave equation for the pion at or near threshold: 
\begin{equation} 
\left[ \nabla^2  - 2{\mu}(B+V_{\rm opt} + V_c) + (V_c+B)^2\right] \psi = 0, 
\label{eq:KG} 
\end{equation} 
\noindent 
where $\hbar = c = 1$ was implicitly assumed in these equations. In this 
expression, $\mu$ is the pion-nucleus reduced mass, $B$ is the complex 
binding energy, $V_c$ is the finite-size Coulomb interaction of the pion 
with the nucleus, including vacuum-polarization terms, all added according 
to the minimal substitution principle $E \to E - V_c$. Interaction terms 
negligible with respect to $2{\mu}V_{\rm opt}$, i.e. $2V_cV_{\rm opt}$ and 
$2BV_{\rm opt}$, were omitted. We use the Ericson-Ericson form~\cite{EEr66} 
\begin{equation} 
2\mu V_{\rm opt}(r) = q(r) + \vec \nabla \cdot 
\left(\frac{\alpha_1(r)}{1+\frac{1}{3}\xi\alpha_1(r)}+\alpha_2(r)\right)
\vec \nabla, 
\label{eq:EE1} 
\end{equation} 
\noindent 
with its $s$-wave part $q(r)$ and $p$-wave part, $\alpha_1(r)$ and 
$\alpha_2(r)$, given by \cite{FGa07} 
\begin{eqnarray} 
q(r) & = & -4\pi(1+\frac{\mu}{m_N})\{ b_0[\rho_n(r)+\rho_p(r)]
 +b_1[\rho_n(r)-\rho_p(r)] \} \nonumber \\
 & &  -4\pi(1+\frac{\mu}{2m_N})4B_0\rho_n(r) \rho_p(r),
\label{eq:EE1s} 
\end{eqnarray} 
\begin{equation} 
\alpha_1(r) = 4\pi(1+\frac{\mu}{m_N})^{-1}
\{c_0[\rho_n(r)+\rho_p(r)]+c_1[\rho_n(r)-\rho_p(r)]\}
\label{eq:EE1p1} 
\end{equation} 
\begin{equation} 
\alpha_2(r) = 4\pi(1+\frac{\mu}{2m_N})^{-1}4C_0\rho_n(r)\rho_p(r),
\label{eq:EE1p2} 
\end{equation} 
\noindent
where $\rho_n$ and $\rho_p$ are the neutron and proton density distributions 
normalized to the number of neutrons $N$ and number of protons $Z$, 
respectively. The coefficients $b_0$ and $b_1$ in Eq.~(\ref{eq:EE1s}) are 
effective, density-dependent pion-nucleon isoscalar and isovector $s$-wave 
scattering amplitudes, respectively, evolving from the free-space scattering 
lengths $b_0^{\rm free}$ and $b_1^{\rm free}$ of Eq.~(\ref{eq:Baru}), and are 
essentially real near threshold. Similarly, the coefficients $c_0$ and $c_1$ 
in Eq.~(\ref{eq:EE1p1}) are effective $p$-wave scattering amplitudes which, 
since the $p$-wave part of $V_{\rm opt}$ acts mostly near the nuclear 
surface, are close to the free-space scattering volumes $c_0^{\rm free}$ 
and $c_1^{\rm free}$ provided $\xi =1$ is applied in the Lorentz-Lorenz 
renormalization of $\alpha_1$ in Eq.~(\ref{eq:EE1}). The parameters $B_0$ and 
$C_0$ represent multi-nucleon absorption and therefore have an imaginary part. 
Their real parts stand for dispersive contributions which often are absorbed 
into the respective single-nucleon amplitudes. Below we focus on the $s$-wave 
part $q(r)$ of $V_{\rm opt}$. 

Regarding the isoscalar amplitude $b_0$, since the free-space value 
$b_0^{\rm free}$ in Eq.~(\ref{eq:Baru}) is exceptionally small, 
it is customary in the analysis of pionic atoms to supplement it by 
double-scattering contributions induced by Pauli correlations which 
give rise to explicit density dependence of the form~\cite{EEr66,KEr69} 
\begin{equation} 
b_0 \rightarrow b_0 - \frac{3}{2\pi}(b_0^2+2b_1^2)p_F, 
\label{eq:b0b} 
\end{equation} 
\noindent 
where $p_F$ is the local Fermi momentum corresponding to the local 
nuclear density $\rho=2p_F^3/(3\pi^2)$.{\footnote{Note added in proof: 
the double-scattering term in (\ref{eq:b0b}) is missing a kinematical factor 
($1+m_\pi/m_N$) \cite{EW88} which, when included, hardly affects our results. 
The role of double-scattering contributions in general will be discussed by 
us in a forthcoming paper.}} 

Regarding the isovector amplitude $b_1$, it is given by the r.h.s. 
of Eq.~(\ref{eq:ddb1}) in terms of a free-space $b_1^{\rm free}$ and 
$\sigma_{\pi N}$. It affects primarily level shifts in pionic atoms with 
$N-Z\neq 0$. However, it affects also $N=Z$ pionic atoms through the 
dominant quadratic $b_1$ contribution to the r.h.s. of Eq.~(\ref{eq:b0b}). 
This dominance follows already at the level of the free-space $b_1^{\rm 
free}$ from a systematic expansion of the pion self-energy up to $O(p^4)$ 
in nucleon and pion momenta within chiral perturbation theory~\cite{KW01}. 
To understand why the in-medium $b_1$ of Eq.~(\ref{eq:ddb1}) enters the 
Pauli-correlation double-scattering contribution to Eq.~(\ref{eq:b0b}), we 
recall how it was introduced in Ref.~\cite{KKW03}. The {\it energy dependence} 
of the pion self-energy operator $\Pi(E,\vec{p},\rho)$ in a uniform medium of 
density $\rho$ was traded there for an equivalent {\it energy independent} 
optical potential, with $b_1^{\rm free}$ promoted to a density dependent 
in-medium $b_1$, Eq.~(\ref{eq:ddb1}). Once done, it is this $b_1$, not 
$b_1^{\rm free}$, that enters as input the formal derivation~\cite{WRW97} 
of the Pauli-correlation double-scattering term. This approach has been 
practised in numerous global fits to pionic atoms by us~\cite{FGa07,FGa14} 
as well as by other groups, e.g., Geissel et al.~\cite{GGG02}. 

An important ingredient in the analysis of pionic atoms are the nuclear 
densities that enter the potential, Eqs.~(\ref{eq:EE1s})--(\ref{eq:EE1p2}). 
With proton densities determined from nuclear charge densities, we vary the 
neutron densities searching for a best agreement with the pionic atoms data. 
A linear dependence of $r_n-r_p$, the difference between the root-mean-square 
(rms) radii, on the neutron excess ratio $(N-Z)/A$ has been recognized  as
a useful and relevant representation, parameterized across the periodic table 
as 
\begin{equation}  
r_n-r_p = \gamma\, \frac{N-Z}{A} + \delta \; ,
\label{eq:rnrp} 
\end{equation} 
\noindent 
with $\gamma$ close to 1.0~fm and $\delta$ close to zero. Two-parameter 
Fermi distributions were used for $\rho_p$ and $\rho_n$ with the same 
diffuseness parameter for protons and neutrons, the so-called `skin' 
shape~\cite{FGa07,Fri09} which was found to yield lower values of $\chi ^2$ 
than other shapes do for pions. Here we used $\delta =-$0.035 fm and varied 
the parameter $\gamma$. With $\gamma$=1~fm, for example, the `neutron skin' 
of $^{208}$Pb is $r_n-r_p=0.177$~fm which agrees well with recent values 
derived specifically for $^{208}$Pb from several sources.{\footnote{For 
example, 0.16$\pm$0.02$\pm$0.04~fm from $\bar p$ atoms~\cite{Klos07}, 
0.156$^{+0.025}_{-0.021}$~fm from $E1$ polarizability studies~\cite{RCNP11}, 
0.15$\pm$0.08~fm from $\pi^-$ atoms~\cite{Fri12}, 0.11$\pm$0.06~fm 
from $\pi^+$ total reaction cross sections~\cite{Fri12}, and 0.15$\pm
$0.03$^{+0.01}_{-0.03}$~fm from coherent pion photoproduction measurements 
at MAMI~\cite{MAMI13}.}}
In what follows, rather than show results as a function of the neutron-excess 
parameter $\gamma$ of Eq.~(\ref{eq:rnrp}), we present results as a function 
of the {\it implied} value of $r_n-r_p$ for $^{208}$Pb, as this quantity has 
been discussed extensively in recent years, e.g. Refs.~\cite{Piekar12,FPH18}, 
particularly in the context of neutron stars.

\section{Results}
\label{sec:res}

\begin{figure}[htb] 
\begin{center} 
\includegraphics[width=0.8\textwidth]{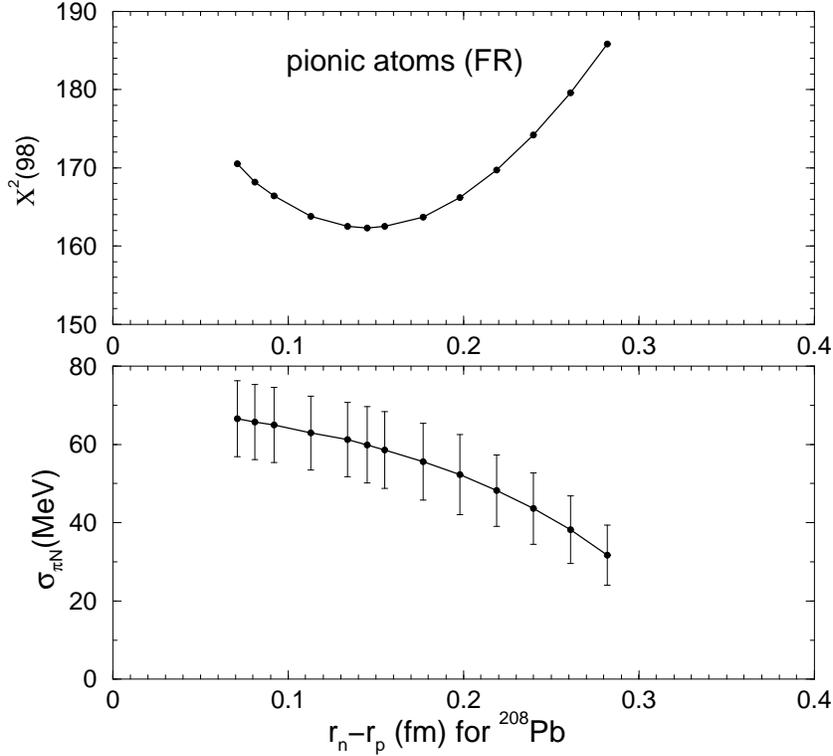} 
\caption{Fits to pionic atoms for different values of the neutron-excess 
radius parameter $\gamma$, presented as the implied neutron skin for 
$^{208}$Pb. Top: $\chi ^2$ values for 98 data points with six adjusted 
parameters, including $\sigma_{\pi N}$. FR denotes finite-range folding 
of $\pi N$ $p$-wave interaction terms. Bottom: derived values of 
$\sigma_{\pi N}$.} 
\label{fig:fits1} 
\end{center} 
\end{figure} 

In line with our previous studies of pionic atoms \cite{FGa07,FGa14} 
we performed global fits to strong interaction level shifts and widths 
across the periodic table, from Ne to U, including `deeply bound' states 
in Sn isotopes and in $^{205}$Pb. This approach provides an average behavior 
of the $\pi N$ interaction parameters within an optical potential model, 
Eqs.~(\ref{eq:EE1})--(\ref{eq:EE1p2}). Fits were made over a wide range of 
values for the neutron-excess radius parameter $\gamma$. The lowest values 
of $\chi ^2$ were obtained, as expected, when varying all eight parameters 
of the optical potential. Whereas the imaginary parts Im$\,B_0$ and Im$\,C_0$ 
were well-determined and hardly varied over a wide range of values tested 
for $\gamma$, their real parts Re$\,B_0$ and Re$\,C_0$ were poorly determined 
and varied over a broad range. Moreover, they displayed correlations with the 
two resulting scattering volumes $c_1$, $c_0$, respectively. With Re$\,B_0$ 
and Re$\,C_0$ kept zero, all the other parameters turned out to be well 
determined. Consequently most of the resulting $\pi N$ interaction terms, but 
not the $\pi N$ $\sigma$ term $\sigma_{\pi N}$, turned out to be independent 
of the neutron-excess radius parameter $\gamma$. We note that when $b_1^{\rm 
free}$ was used in the double scattering term (\ref{eq:b0b}) instead of the 
in-medium $b_1$ form (\ref{eq:ddb1}), the lowest $\chi ^2$ increased by about 
5 units. With an achieved $\chi ^2$ per degree of freedom of 1.7 this increase 
means three standard deviations. 

Extensive fits essentially displayed {\it correlations} between rms radii 
of the neutron density distribution and the resulting $\sigma_{\pi N}$, 
as shown in Fig.~\ref{fig:fits1}. The figure shows fits with six adjusted 
parameters, namely $b_0$, $\sigma _{\pi N}$, Im$\,B_0$, $c_0$, $c_1$ and 
Im$\,C_0$. As in earlier work \cite{FGa07} a finite range (FR) folding of 
rms radius of 0.9 fm was applied to the $\pi N$ $p$-wave interaction terms. 
The bottom part of the figure shows the derived $\sigma_{\pi N}$ values 
with their uncertainties. An interesting by-product of these fits is the 
value 0.15$\pm$0.03 fm of the implied neutron skin of $^{208}$Pb, taken 
from the minimum of the $\chi ^2$ curve in the top part of the figure, 
in agreement with the values cited in a footnote to the text at the end 
of Sect.~\ref{sec:meth} above. 

\begin{figure}[htb] 
\begin{center} 
\includegraphics[width=0.75\textwidth]{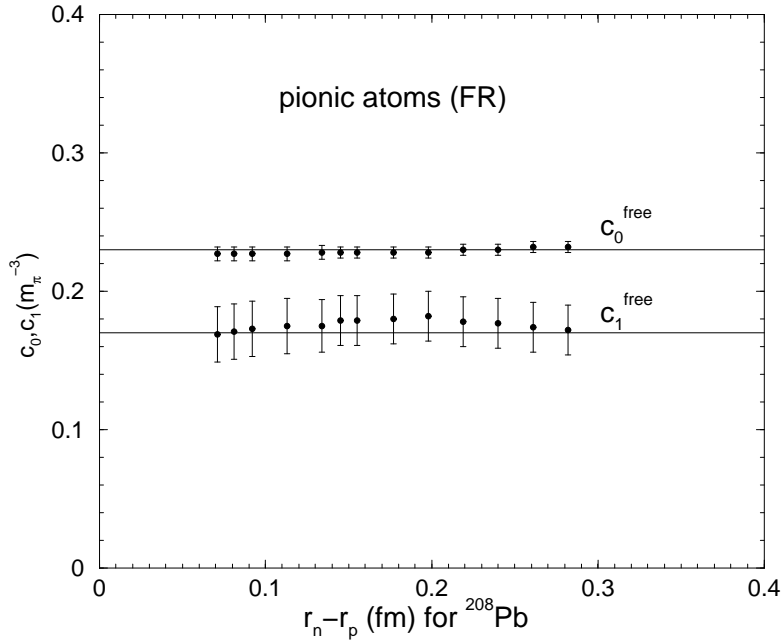} 
\caption{Values of the $\pi N$ $p$-wave parameters $c_0$ and $c_1$ obtained 
in the FR fits of Fig.~\ref{fig:fits1}. The horizontal lines mark the 
SAID~\cite{SAID06} free-space values $c_0^{\rm free}$ and $c_1^{\rm free}$ 
of the $\pi N$ scattering volumes.} 
\label{fig:c0c1} 
\end{center} 
\end{figure} 

In the fits shown in Fig.~\ref{fig:fits1}, the single-nucleon isoscalar 
$c_0$ and isovector $c_1$ parameters of the $\pi N$ $p$-wave potential 
$\alpha_1(r)$ turned out to agree with the corresponding values of the 
free-space $\pi N$ scattering volumes. This is shown in Fig.~\ref{fig:c0c1}. 

\begin{figure}[htb] 
\begin{center} 
\includegraphics[width=0.8\textwidth]{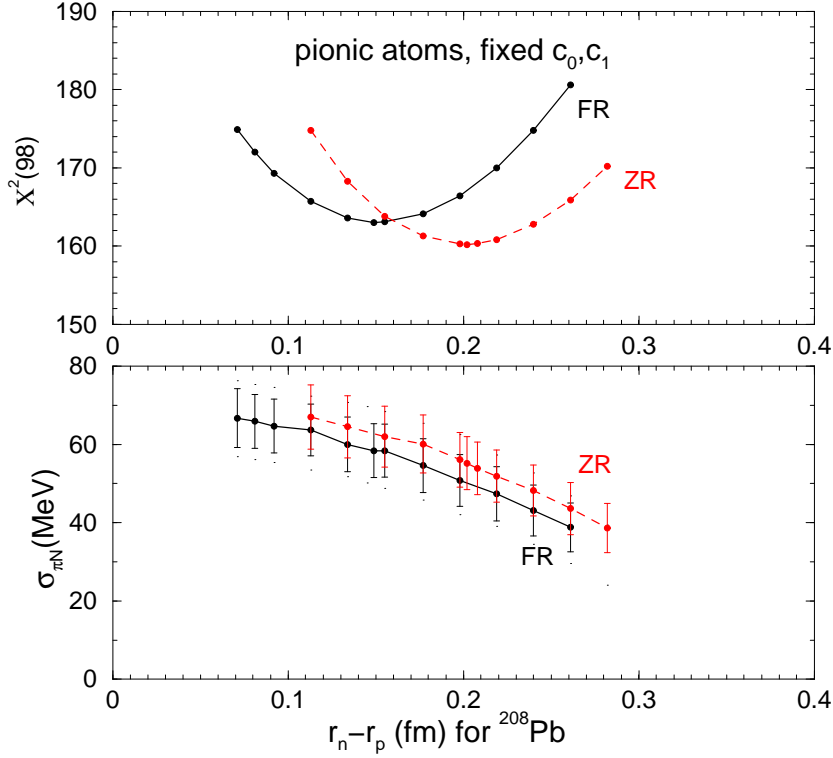} 
\caption{FR and ZR fits to pionic atoms for different values of the 
neutron-excess radius parameter $\gamma$, presented as the implied 
$^{208}$Pb neutron skin, with fixed values of $c_0$ and $c_1$. 
Top: $\chi ^2$ values for 98 data points. Bottom: fitted values of 
$\sigma_{\pi N}$. Black (red) solid (dashed) lines correspond to FR (ZR).} 
\label{fig:fits12} 
\end{center} 
\end{figure} 

With $c_0$ and $c_1$ hardly dependent on the neutron densities, one could 
keep these fixed during fits to reduce the uncertainties of the resulting 
values of $\sigma_{\pi N}$. Figure~\ref{fig:fits12} shows two such fits with 
fixed values, both analogous to Fig.~\ref{fig:fits1}, one with $p$-wave 
finite-range folding (FR, solid lines, black), and one without folding 
(ZR, dashed lines, red). In both parts of Fig.~\ref{fig:fits12} the red 
curves are shifted to the right of the corresponding black curves, but 
for the best fit values of $\sigma_{\pi N}$, at the minima of $\chi ^2$, 
there is hardly any difference between the FR and ZR models, regardless 
of the $\sim$0.06~fm difference between the best implied values of 
the $^{208}$Pb skin in these models. With fixed $c_0$ and $c_1$, the 
fitting errors are indeed smaller than those in Fig.~\ref{fig:fits1}. 
The average value for $\sigma_{\pi N}$ from Fig.~\ref{fig:fits12} is 
$\sigma_{\pi N}=57\pm7$ MeV. For $p$-wave FR folding with rms radius smaller 
than 0.9~fm the resulting curves in Fig.~\ref{fig:fits12} are located between 
the 0.9~fm FR curve and the ZR curve. We note that a size of order 1~fm 
appears naturally for $\pi N$ $p$-wave form factors fitted to the $\Delta$ 
resonance $P_{33}$ phase shifts~\cite{GG11}.

\section{Discussion and summary}
\label{sec:disc}

The pionic atoms fits and the value of the $\pi N$ $\sigma$ term 
$\sigma_{\pi N}$ extracted in the present work are based on the in-medium 
renormalization of the near-threshold $\pi N$ isovector scattering amplitude 
$b_1$ as given by Eq.~(\ref{eq:ddb1}), derived from Eq.~(\ref{eq:fpi}) for 
the leading order in-medium decrease of the quark condensate $<\bar q q>$. 
Higher order corrections to this simple form have been proposed in the 
literature and are discussed briefly below to see how much they affect our 
fitted value of $\sigma_{\pi N}$. Generally, one does not expect appreciable 
corrections simply because typical nuclear densities probed in pionic atoms 
are only about 0.6 \cite{YHi03} or even 0.5 \cite{SMa83} of nuclear matter 
density. A representative effective density of $\rho_{\rm eff}=0.1$~fm$^{-3}$ 
is used for the two types of corrections discussed below. 

Kaiser et al.~\cite{KHW08} extended Eq.~(\ref{eq:fpi}) to  
\begin{equation} 
\frac{<\bar q q>_{\rho}}{<\bar q q>}=1-\frac{\rho}{f_{\pi}^2}\left[
\frac{\sigma_{\pi N}}{m_{\pi}^2}\left(1-\frac{3p_F^2}{10m_N^2}+\frac{9p_F^4}
{56m_N^4}\right)+\frac{\partial E(\rho)/A}{\partial m_{\pi}^2}\right], 
\label{eq:KHW} 
\end{equation} 
\noindent 
accounting for kinetic energy contributions up to order $m_N^{-3}$ in 
the Fermi gas model plus $NN$ correlation contributions from one- and 
two-pion interaction terms. At density $\rho_{\rm eff}=0.1$~fm$^{-3}$ and 
for $\sigma_{\pi N}=60$~MeV the r.h.s. of Eq.~(\ref{eq:KHW}) is about 0.75, 
larger than the purely linear density expression by about 0.03. Most of 
this increase is owing to the $NN$ correlation contributions. If we wish 
to absorb at $\rho_{\rm eff}$ this departure from linearity in $\rho$ 
into an effective linear density form, Eq.~(\ref{eq:fpi}), we need 
to {\it increase} our fitted $\sigma_{\pi N}$ value by about 7 MeV. 
A smaller increase results by following a chiral approach up to 
next-to-leading order~\cite{LOM10}. 

Jido and collaborators~\cite{JHK08} extended the GOR expression 
Eq.~(\ref{eq:fpi}) by including the increase of the in-medium pion mass 
$m_{\pi}(\rho)$ in symmetric nuclear matter{\footnote{In asymmetric nuclear 
matter the charged pion masses split, with $m_{\pi^-}$ increasing further 
owing to the repulsive $b_1$ isovector $s$-wave $\pi N$ interaction 
term~\cite{KW01}. This effect is disregarded in the estimate given below.}} 
from its free-space value $m_{\pi}$:   
\begin{equation} 
\frac{f_\pi^2(\rho)}{f_\pi^2} = \frac{m_{\pi}^2}{m_{\pi}^2(\rho)}\,
\frac{<\bar q q>_{\rho}}{<\bar q q>}\simeq\frac{m_{\pi}^2}{m_{\pi}^2(\rho)}\,
\left(1 - \frac{\sigma_{\pi N}}{m_{\pi}^2 f_{\pi}^2}\,\rho\right), 
\label{eq:JHK} 
\end{equation}
\noindent 
and also by adding corrections of order $\rho^{4/3}$~\cite{GJi13,GJi14} 
which at $\rho_{\rm eff}$ are negligible. The pion mass dependence in 
Eq.~(\ref{eq:JHK}) leads to the following modification of Eq.~(\ref{eq:ddb1}) 
for the near-threshold $\pi N$ isovector amplitude: 
\begin{equation}   
b_1=b_1^{\rm free}\,\frac{m_{\pi}^2(\rho)}{m_{\pi}^2}
\left(1-\frac{\sigma_{\pi N}}{m_{\pi}^2f_{\pi}^2}\,\rho\right)^{-1}. 
\label{eq:ddb1rev} 
\end{equation}
\noindent
The pion mass in $N=Z$ isospin-zero symmetric nuclear matter increases 
from its free-space value $m_{\pi}$ to $m_{\pi}(\rho)$ owing to the weakly 
repulsive $b_0$ isoscalar $s$-wave $\pi N$ interaction term. Identifying 
the in-medium pion mass with $E(\vec p=0)$ in the dispersion equation 
(\ref{eq:disp}) and using Eq.~(\ref{eq:EE1s}) with an appropriate subthreshold 
value~\cite{FGa14} $\overline{b}_0=-0.011(2)\,m_{\pi}^{-1}$ corresponding to 
our best fit threshold value $b_0=-0.022(2)\,m_{\pi}^{-1}$, we obtain for the 
difference $\delta m_{\pi}^2=m_{\pi}^2(\rho)-m_{\pi}^2$ at $\rho_{\rm eff}=
0.1$~fm$^{-3}$: 
\begin{equation}  
\delta m_{\pi}^2\approx -4\pi\left(1+\frac{m_{\pi}}{m_N}\right)
\overline{b}_0\,\rho_{\rm eff}=0.045(8)\, m_{\pi}^2 \,, 
\label{eq:delms} 
\end{equation} 
\noindent  
or equivalently $m_{\pi}^2(\rho_{\rm eff})/m_{\pi}^2=1.045(8)$. 
Interestingly, this isoscalar contribution to $m_{\pi}^2(\rho)$ agrees with 
a rescattering contribution to the pion in-medium self energy, derived in 
Ref.~\cite{CEO03} with the purpose of providing additional renormalization 
of $b_1^{\rm free}$ beyond the leading $\sigma_{\pi N}$ contribution given 
by Eq.~(\ref{eq:ddb1}).{\footnote{Ref.~\cite{CEO03} deals with several other 
in-medium modifications that require separate discussion. In particular, 
their rescattering contribution, which as pointed out here is equivalent 
to in-medium pion mass renormalization, was considered earlier by Delorme 
et al.~\cite{DEE92}. For this contribution, adding a factor $\pi$ missing 
in the denominator of Eq.~(28) in Ref.~\cite{CEO03}, one obtains $(\delta 
b_1/b_1^{\rm free})\approx (3p_F\rho_{\rm eff})/(16\pi^2f_{\pi}^4)=0.0454$, 
in agreement with Eq.~(\ref{eq:delms}) above.}} With this increased in-medium 
pion mass, our best-fit central value of $\sigma_{\pi N}$=57~MeV {\it 
decreases}, by just (7$\pm$1)~MeV, to (50$\pm$1)~MeV. Perhaps fortuitously, 
the two higher-order effects considered here upon deriving $\sigma_{\pi N}$ 
from pionic atoms, Eqs.~(\ref{eq:KHW}) and (\ref{eq:ddb1rev}), cancel each 
other. 

It is worth recalling that the attractive isoscalar $p$-wave $\pi N$ 
interaction term was disregarded in this uniform nuclear matter estimate 
where the pion momentum vanishes. In finite-size nuclei, however, multiplying 
$p_F(\rho_{\rm eff})$ by $m_{\pi}/(m_N+m_{\pi})$ a representative pion 
effective momentum of $p_{\rm eff}=29.1$~MeV is obtained. This leads to the 
following $p$-wave contribution:  
\begin{equation}  
\delta m_{\pi}^2\approx -\frac{4\pi(1+\frac{m_{\pi}}{m_N})^{-1}c_0\,
\rho_{\rm eff}}{1+\frac{1}{3}4\pi(1+\frac{m_{\pi}}{m_N})^{-1}c_0\,
\rho_{\rm eff}}\,p_{\rm eff}^2 
=-0.025\, m_{\pi}^2 \,,  
\label{eq:delmp} 
\end{equation} 
\noindent 
using $c_0^{\rm free} = 0.230\,m_{\pi}^{-3}$. Adding up these $s$-wave 
and $p$-wave contributions, we get $m_{\pi}^2(\rho_{\rm eff})/m_{\pi}^2=
1.020(8)$, leading to a decrease of our best-fit $\sigma_{\pi N}$ central 
value of 57~MeV, by only (3$\pm$1)~MeV, to (54$\pm$1)~MeV.  

To conclude the discussion, we note that unlike most determinations of 
$\sigma_{\pi N}$ that rely heavily on the vanishingly small and highly model 
dependent value of the free-space $\pi N$ isoscalar scattering length 
$b_0^{\rm free}$, the present work is based on the considerably larger and 
nearly model independent value of the free-space $\pi N$ isovector scattering 
length $b_1^{\rm free}$. The dependence of $\sigma_{\pi N}$ on the input 
free-space $\pi N$ scattering lengths, within any specific hadronic model 
calculation, is given according to the Bonn-J\"{u}lich (BJ) group~\cite{Hof16} 
by  
\begin{equation} 
\sigma_{\pi N}\approx(59\pm 3)~{\rm MeV} + 1.116\,\Delta b_0^{\rm free} 
+ 0.390\, \Delta b_1^{\rm free}, 
\label{eq:hof16} 
\end{equation} 
where (59$\pm$3)~MeV is the BJ calculated $\sigma_{\pi N}$ value~\cite{Hof15} 
and $\Delta b_j^{\rm free}$, $j=0,1$, is the difference between the values 
of $b_j^{\rm free}$ (in units of $10^{-3}\,m_{\pi}^{-1}$) used in that 
specific model and in the BJ calculation. Two sets were suggested by BJ 
for $(b_0^{\rm free},\,b_1^{\rm free})$, 
\begin{equation} 
{\rm BJ:}\,\,\,\,\,\,\,\,(-0.9,\,-85.3)\cdot 10^{-3}\,m_{\pi}^{-1},
\,\,\,(+7.9,\,-85.4)\cdot 10^{-3}\,
m_{\pi}^{-1},
\label{eq:BJ} 
\end{equation} 
depending on how charge dependence is considered. These two sets differ mostly 
in the $b_0^{\rm free}$ values. To demonstrate the use of Eq.~(\ref{eq:hof16}) 
we refer to the evaluation of the $\pi N$ $\sigma$ term in Ref.~\cite{SCW13} 
from $\pi^{\pm}p$ scattering data taken by the CHAOS group at 
TRIUMF~\cite{Denz06}. Extrapolating from the lowest pion kinetic energy of 
19.9~MeV reached in the experiment, the value used in Ref.~\cite{SCW13} was 
$b_0^{\rm free}=(-9.7\pm 0.9)\cdot 10^{-3}\,m_{\pi}^{-1}$. 
Eq.~(\ref{eq:hof16}) `predicts' then $\sigma_{\pi N}$=49$\pm$3 or 
39$\pm$3~MeV, depending on the choice made for $b_0^{\rm free}$ in 
Eq.~(\ref{eq:BJ}), in rough agreement with the value $\sigma_{\pi N}$=(44$
\pm$12)~MeV derived in Ref.~\cite{SCW13}. Similarly, the increase of 
$\sigma_{\pi N}$ from the older value (45$\pm$8)~MeV derived by 
Gasser, Leutwyler and Sainio~\cite{GLS91} to the very recent 
(59$\pm$3)~MeV~\cite{Hof15} is related, according to Eq.~(\ref{eq:hof16}), 
to the use by the BJ group of the more precise $\pi N$ scattering lengths 
as extracted recently from $\pi^-$H and $\pi^-$D atoms. 

In conclusion, we have derived in this work a value of $\sigma_{\pi N}=57
\pm 7$ MeV from a large scale fit to pionic atoms observables, in agreement 
with the relatively high values reported in recent studies based on modern 
hadronic $\pi N$ phenomenology~\cite{RdE18}, but in disagreement with the 
low $\sigma_{\pi N}$ values reached in the modern lattice QCD calculations, 
e.g.~\cite{JLQCD18}. Our derivation is based on the model introduced by Weise 
and collaborators~\cite{Wei00,Wei01,KKW03} for the in-medium renormalization 
of the $\pi N$ near-threshold isovector scattering amplitude, using its 
leading density dependence Eq.~(\ref{eq:ddb1}), and was found robust in 
fitting the wealth of pionic atoms data against variation of other $\pi N$ 
interaction parameters that enter the low-energy pion self-energy operator. 
The two types of model corrections beyond the leading density dependence 
considered here were found to be relatively small, a few MeV each, and partly 
canceling each other. Further model studies are necessary to confirm this 
conclusion.

\section*{Acknowledgments} 
We thank Wolfram Weise for a useful communication~\cite{Wei18} on in-medium 
and partial restoration of chiral symmetry effects on $\sigma_{\pi N}$.

\end{document}